\documentclass[twocolumn,showpacs,preprintnumbers,amsmath,amssymb,aps]{revtex4}
\usepackage{graphicx}
\usepackage{epsfig}
\usepackage{pict2e}

\def\lsim{\raise0.3ex\hbox{$\;<$\kern-0.75em\raise-1.1ex\hbox{$\sim\;$}}}
\def\gsim{\raise0.3ex\hbox{$\;>$\kern-0.75em\raise-1.1ex\hbox{$\sim\;$}}}

\newcommand{\be}{\begin{eqnarray}}
\newcommand{\ee}{\end{eqnarray}}

\def\bea{\begin{eqnarray}}
\def\eea{\end{eqnarray}}

\usepackage{subfigure}
\usepackage{float}
\usepackage{color}
\usepackage[colorlinks=true
,urlcolor=blue
,citecolor=blue
,linkcolor=blue
,pagecolor=blue
,linktocpage=true
,pdfproducer=medialab
]{hyperref}

\begin{document}

\title{Muon Anomalous Magnetic Moment in SUSY $B-L$ Model with Inverse Seesaw}
\author{Shaaban Khalil$^{a}$ and Cem Salih \"{U}n$^{b}$}
\affiliation{$^a$ Center of Fundamental Physics, Zewail City of Science and Technology, 6 October City, Cairo Egypt \\
$^b$  Department of Physics, Uluda{\~g} University, Bursa TR16059, Turkey.}

\begin{abstract}
\noindent 
Motivated by the tension between the Higgs mass and muon $g-2$ in minimal supersymmetric standard model (MSSM), we  analyze the muon $g-2$ in supersymmertic $B-L$  extension of the standard model (BLSSM) with inverse seesaw mechanism. In this model, the Higgs mass receives extra important radiative corrections proportional to large neutrino Yukawa coupling. We point out that muon $g-2$ also gets significant contribution, due to the constructive interferences of light neutralino effects. The light neutralinos are typically the MSSM Bino like and the supersymmetric partner of $U(1)_{B-L}$ gauge boson ($\tilde{B}'$ino).  We show that with universal soft supersymmetry breaking terms, the muon $g-2$ resides within $2\sigma$ of the measured value, namely $ \sim 17 \times 10^{-10}$, with Higgs mass equal to 125 GeV .

\end{abstract}

\maketitle

The Standard Model (SM) prediction for the anomalous magnetic moment of the muon, $a_{\mu}=(g-2)_{\mu}/2$ (hereafter muon $g-2$) has a discrepancy with the experimental results:
\begin{equation}\hspace{-0.3cm}
\Delta a_{\mu}\equiv a_{\mu}^{{\rm exp}} - a_{\mu}^{{\rm SM}} = (28.7\pm 8)\times 10^{-10}~(1\sigma).
\label{damuexp}
\end{equation}
This discrepancy has survived after performing highly accurate theoretical calculations \cite{Davier:2010nc} within the SM framework and experimental analyses \cite{Bennett:2006fi}; and hence, it can be resolved or ameliorated by contributions from new physics beyond the SM (BSM). If supersymmetry (SUSY), as one of the forefront candidates for the BSM physics,  is a solution to the muon $g-2$, the SUSY particles, namely, smuon and weak gaugino (Bino or Wino) masses should be around a few hundred GeV, in order to utilize the supersymmetric contributions \cite{Moroi:1995yh}.

However, the observation of the Higgs boson of mass about 125 GeV requires rather heavy sparticle spectrum within the MSSM framework, and it results in a strong tension in simultaneous resolution for both the 125 GeV Higgs boson and the muon $g-2$ problem since SUSY contributions to muon $g-2$ is suppressed by the heavy spectrum. Non-universality in gaugino and/or scalar masses may remove this tension \cite{Ajaib:2014ana}, nevertheless in this case SUSY models will have plenty of free parameters and will lose their productivity.  

In this article we show that this tension can be alleviated  in the $U(1)_{B-L}$ extended Supersymmetric Standard Model (BLSSM), which is one of the interesting non-minimal realizations of supersymmetry.  The BLSSM is also well motivated by the established existence of non-zero neutrino masses \cite{Wendell:2010md}. It turns out that, in this class of model, the scale of $B-L$ symmetry breaking can be related to the scale of SUSY breaking \cite{Khalil:2007dr}, therefore, a TeV scale type I or inverse seesaw mechanism can be naturally implemented \cite{Khalil:2010iu}.  The one-loop radiative corrections to the SM-like Higgs boson mass, due to the right-handed (s)neutrinos in BLSSM, with inverse seesaw mechanism, provide
new contribution to the Higgs boson mass in addition to the stop sector of MSSM  \cite{Elsayed:2011de}. 
Thus the lower bound imposed by Higgs mass on the universal gaugino soft masses $m_{1/2}$ is reduced, which makes possible to find solutions with the light weak gauginos. Moreover, in BLSSM with inverse seesaw the $g-2$ may receive new contributions, in addition to the usual MSSM ones, due to the extension of the neutralino sector by SM singlet $(B-L)$ Higgsino and $B'$-ino and also due to the possibility that one of the right-handed sneutrinos is light (due to large mixing between right-handed sneutrinos and right-handed anti-sneutrinos). We emphasize that within BLSSM with inverse seesaw the $g-2$ resides within $2\sigma$ of the measured value. We show that in this model, even if we are restricted to the universal boundary conditions at the GUT scale, we can account for the Higgs mass and keep the lightest neutralino to be as light as about 200 GeV. 

The BLSSM is based on the gauge group $SU(3)_C \times SU(2)_L  \times U(1)_Y \times U(1)_{B-L}$. One can implement the seesaw mechanism by adding the right-handed neutrino one per family as required by the anomaly cancellation. In order to count for the right-handed neutrino contributions, we employ the inverse seesaw model in our study. The BLSSM with inverse seesaw includes, in addition to the MSSM particle contents, two SM singlet chiral Higgs superfields $\chi_{1,2}$ and three sets of SM singlet chiral superfields $\nu_{i}$, $s_{1_{i}}$, $s_{2_{i}}$ \cite{Khalil:2015naa}. The Superpotential of this model is given by \cite{Staub:2013tta}
\bea
W &=&  - {\mu_{\eta}} \,\hat{\chi}_1\,\hat{\chi}_2\,+\mu\,\hat{H}_u\,\hat{H}_d\,+\mu_S\,\hat{s}_2\,\hat{s}_2\,- Y_d \,\hat{d}\,\hat{q}\,\hat{H}_d\nonumber\\
&-& Y_e \,\hat{e}\,\hat{l}\,\hat{H}_d\,+Y_u\,\hat{u}\,\hat{q}\,\hat{H}_u\,+Y_s\,\hat{\nu}\,\hat{\chi}_1\,\hat{s}_2 +Y_\nu\,\hat{\nu}\,\hat{l}\,\hat{H}_u. ~~~
\label{superpotential}
\eea
The definition of the parameters appear in $W$, the corresponding soft SUSY breaking terms and the details of the associate spectrum can be found in Refs.~\cite{Khalil:2015naa, Staub:2013tta}.  Note that the $U(1)_Y$ and $U(1)_{B-L}$ gauge kinetic mixing can be absorbed in the covariant derivative redefinition. Therefore, this mixing provides another source of coupling between MSSM and $B-L$ sectors.  Here, we will focus only on the particles involved in the $g-2$ loops, namely light neutralino, smuon and chargino, sneutrino.  

Considering the additional SM singlet fields of BLSSM with inverse seesaw mentioned above the $B-L$ extension may modify only the neutral sectors of the MSSM only, and hence the chargino and slepton mass matrices remain intact. The  $7\times 7$ neutralino mass matrix, in the basis: $\left(\tilde{B}, \tilde{W}^0, \tilde{H}_d^0, \tilde{H}_u^0, {\tilde{B}{}'}, \tilde{\chi_1}, \tilde{\chi_2}\right)$, can be found in Ref. \cite{Staub:2013tta}. One can easily show that depending on the ratio of the couplings $g_1$ and $g_{BL}$, the lightest neutralino could be $B$-ino ($\tilde{B}$) or $B'$-ino ($\tilde{B}'$) like. It is worth noting that in order to account for the Higgs mass, $m_0$ should be of order TeV, thus $\mu$ and $\mu_\eta$, which are determined from  electroweak and $B-L$ symmetry breaking conditions, respectively, are also of order TeV scale. Hence, the lightest eigenstates of neutralinos are mostly formed by the neutral gauginos, since the neutral Higgsinos are quite heavy.  Now we turn to the sneutrino mass matrix.  If we write $\tilde{\nu}_{L,R}$ and  $\tilde{S}_2$ as $\tilde{\nu}_{L,R} = \frac{1}{\sqrt{2}}(\phi_{L,R} + i \sigma_{L,R})$ and 
$\tilde{S}_{2} = \frac{1}{\sqrt{2}}(\phi_{S} + i \sigma_{S})$, then we can get the CP-odd/even sneutrinos matrices as given in \cite{Staub:2013tta}. 

The supersymmetric contributions to $a_{\mu}$ in BLSSM can be split into neutralino and chargino parts as for the MSSM \cite{Moroi:1995yh},
\bea
&&\hspace{-0.75cm}a_{\mu}^{\chi^0} = \frac{m_{\mu}}{16 \pi^2} \sum_{m,i}
\left\{- \frac{m_{\mu}}{6 m_{\tilde{\mu}_m}^2 \left(1 - x_{mi}\right)^{4}}   
\left(\vert N_{mi}^L \vert^2 + \vert N_{mi}^R \vert^2 \right) \right.\nonumber\\
\hspace{-0.75cm}&\times& \left.\left( 1 - 6 x_{mi} +\!3 x^2_{mi} +2 x_{mi}^3 - 6 x_{mi}^2 
\ln x_{mi} \right)\right.\\
\hspace{-0.75cm}&+&\left.\frac{m_{\chi^0_i}}{m_{\tilde{\mu}_m}^2 ( 1 - x_{mi})^{3}}
N_{mi}^L N_{mi}^R ( 1 - x_{mi}^2 + 2 x_{mi} \ln x_{mi})\!\right\}\nonumber
\label{neutralino}
\eea

\bea
&&\hspace{-0.75cm} a_{\mu}^{\chi^{\pm}} = \frac{m_{\mu}}{16 \pi^2} \sum_k \left\{ 
\frac{m_{\mu}}{3 m_{\tilde{\nu}}^2 \left(1-x_k\right)^4}
\left(\vert C_k^L \vert^2 + \vert C_k^R \vert^2 \right)\right.\nonumber\\
\hspace{-0.75cm}&\times& \left. \left( 1 + 1.5 x_k + 0.5 x_k^3
- 3 x_k^2 + 3 x_k \ln x_k \right)\right.\\
\hspace{-0.75cm}&-&\left.
\frac{ 3 m_{\chi^\pm_k}}{m_{\tilde{\nu}}^2 \left(1-x_k\right)^3} C_k^L
C_k^R \left( 1 - \frac{4 x_k}{3} + \frac{x_k^2}{3} + \frac{2}{3}
\ln x_k\right) \right\}\nonumber
\label{chargino}
\eea
where $x_{mi} = m_{\chi^0_i}^2/m_{\tilde{\mu}_m}^2$, $x_k = m_{\chi^\pm_k}^2/
m_{\tilde{\nu}}^2$\ , and
\bea
&&\hspace{-0.75cm}N^{L}_{aij}=-\frac{i}{2}\left[ \sqrt{2}(2g_{1}+\tilde{g})N^{*}_{a1}(U^{*}_{\tilde{\mu}}Z^{\mu\dagger}_{R})_{ij}+\sqrt{2}(2\tilde{g}+g_{BL}) \right.\nonumber\\
&&\hspace{-0.75cm}\left. N^{*}_{a5}(U^{*}_{\tilde{\mu}}Z^{\mu\dagger}_{R})_{ij} +2N^{*}_{a3}(U_{\tilde{\mu}}Y_{\mu}^{T}Z^{\mu\dagger}_{R})_{ij}  \right]
\label{NLaij}
\eea

\bea
&&\hspace{-0.75cm} N^{R}_{aij} = \frac{i}{2}\left[-2N_{a3}(Z^{\mu}_{L}Y_{\mu}^{\dagger}U_{\tilde{\mu}}^{\dagger})_{ij} \!+\! \sqrt{2}(g_{1}\!+\!\tilde{g})N^{*}_{a1}(Z^{\mu}_{L}U_{\tilde{\mu}}^{\dagger})_{ij} \right.\nonumber\\
&&\hspace{-0.75cm}\left. \!+\sqrt{2}g_{2}N^{*}_{a2}(Z^{\mu}_{L}U_{\tilde{\mu}}^{\dagger})_{ij} \!+\!\sqrt{2}(\tilde{g}\!+\!g_{BL})N^{*}_{a5}(Z^{\mu}_{L}U_{\tilde{mu}}^{\dagger})_{ij}\!\right]
\label{NRaij}
\eea

\begin{equation}
\hspace{-0.1cm}
C^{L}_{bij} = \frac{-1}{\sqrt{2}}(U^{*}_{\tilde{\chi}^{-}}\!)_{b2} (U_{\tilde{\nu}^{i}}^{*}Y_{\mu}^{T}Z^{\mu\dagger}_{R})_{ij}\!+\!\frac{i}{2}(U^{*}_{\tilde{\chi}^{-}})_{b2}(U_{\tilde{\nu}^{R}}^{*}Y_{\mu}^{T}Z^{\mu\dagger}_{R})_{ij}
\label{CLbij}
\end{equation}

\bea
&&\hspace{-0.75cm}C^{R}_{bij}=\frac{1}{\sqrt{2}}\left[ g_{2} (U_{\tilde{\chi}^{+}})_{b1} (Z^{\mu}_{L}U_{\tilde{\nu}^{i}}^{\dagger})_{ij}-
(U_{\tilde{\chi}^{+}})_{b2}(Z^{\mu}_{L}Y_{\nu}^{\dagger}U_{\tilde{\nu}^{i}}^{\dagger})_{ij} \right]\nonumber\\
\hspace{-0.75cm}&-&\frac{i}{\sqrt{2}}\left[ g_{2} (U_{\tilde{\chi}^{+}})_{b1} (Z^{\mu}_{L}U_{\tilde{\nu}^{R}}^{\dagger})_{ij} - (U_{\tilde{\chi}^{+}})_{b2}(Z^{\mu}_{L}Y_{\nu}^{\dagger}U_{\tilde{\nu}^{R}}^{\dagger}) \right]
\label{CRbij}
\eea
where $Z_{L,R}^{\mu}$, $U_{\tilde{\mu}}$ are the rotation matrices which diagonalize the muon and smuon mass matrices respectively, while $U_{\tilde{\nu}^{i}}$ and $U_{\tilde{\nu}^{R}}$ diagonalize the CP-odd and CP-even sneutrino mass matrices. Note that one can neglect the mixing between slepton families, and consider the smuon mass matrix as $2\times 2$ matrix separately from the first and third families. In addition, the mixing between two smuons is proportional to the Yukawa coupling associated with muon, which is of the order $\sim 10^{-4}$, and hence left- and right-handed smuons are approximately match with the mass eigenstates, and hence, the rotation matrix for the smuons, $U_{\tilde{\mu}}$ can be set to unity in a good approximation. A similar discussion holds for the muon mass matrix diagonalized by $Z_{L,R}^{\mu}$. 

As seen from Eqs.(\ref{NLaij},\ref{NRaij}), the Bino contribution is in a similar form as obtained in the MSSM, but in BLSSM it has an enhancement by the gauge mixing between $U(1)_{Y}$ and $U(1)_{B-L}$ characterized by the coupling $\tilde{g}$. It is worth emphasizing the contribution from $B'-$ino ($\tilde{B}'$). It contributes to $a_{\mu}$ through interactions with muon governed by $B-L$ gauge group. In addition, its contribution exhibits an enhancement with the gauge kinetic mixing. Moreover, since it is allowed to be as light as Bino, and even lighter, the lightest neutralino can be formed to be mostly $\tilde{B}'$ or $\tilde{B}-\tilde{B}'$ mixing. Thus, one can expect its contribution to be comparable with that from Bino in BLSSM; i.e. $N_{a1}\approx N_{a5}$ numerically. We also present the contribution from the Higgsino component of the Neutralino; however, due to the smallness of $Y_{\mu}$, its contribution is strongly suppressed.

Similarly Eqs.(\ref{CLbij},\ref{CRbij}) reveal the contribution from the chargino expressed in terms of CP-odd and CP-even sneutrino sectors separately. Note that Eqs.(\ref{CLbij},\ref{CRbij}) hold approximately, and one can combine these two sectors in the case of strong mixing between them by summing over $j=1,\ldots,9$. The contribution denoted by $C^{L}_{bij}$ are mostly suppressed because of $Y_{\mu}$. On the other hand, $C^{R}_{bij}$ is expected to dominate  in the chargino contribution to $a_{\mu}$. It arises from the interactions between muon and sneutrinos through SU(2) interactions as shown in the first and third terms in Eq.(\ref{CRbij}). The remaining terms are from the Yukawa interactions, and they cannot be neglected, since $Y_{\nu}$ is allowed to be of the order $\mathcal{O}(1)$ when one employs the inverse seesaw mechanism in BLSSM.

\begin{figure}
\hspace{-1cm}\subfigure{\includegraphics[height=6cm,width=7.75cm]{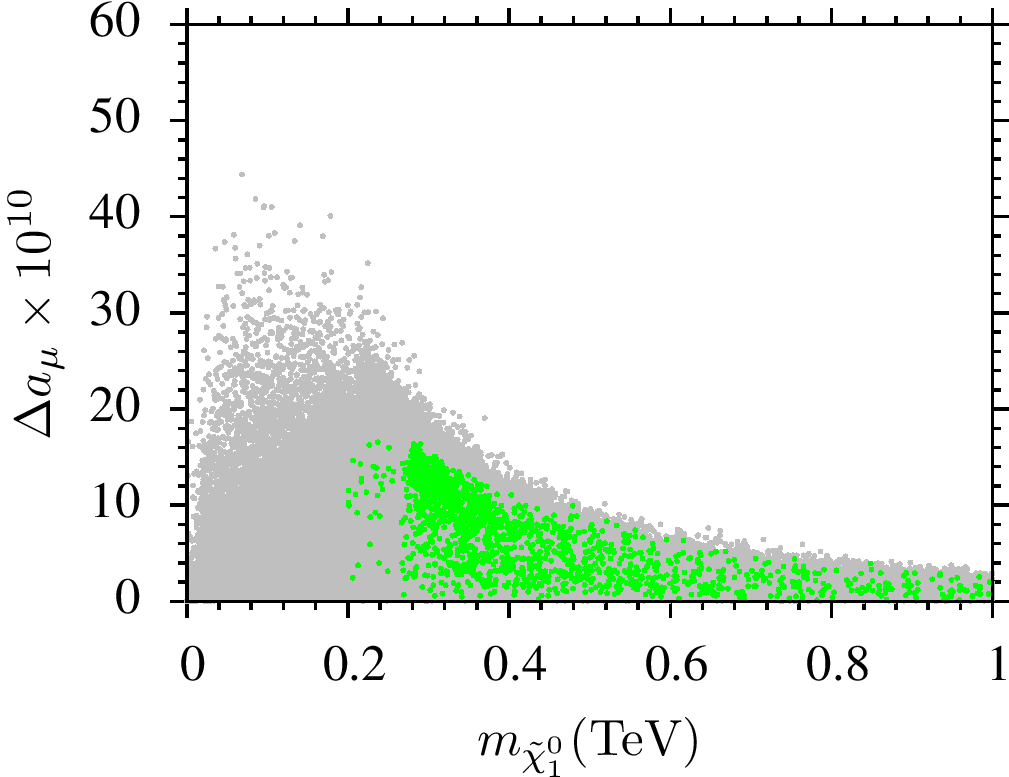}}\\
\hspace{-1cm}\subfigure{\includegraphics[height=6cm,width=7.75cm]{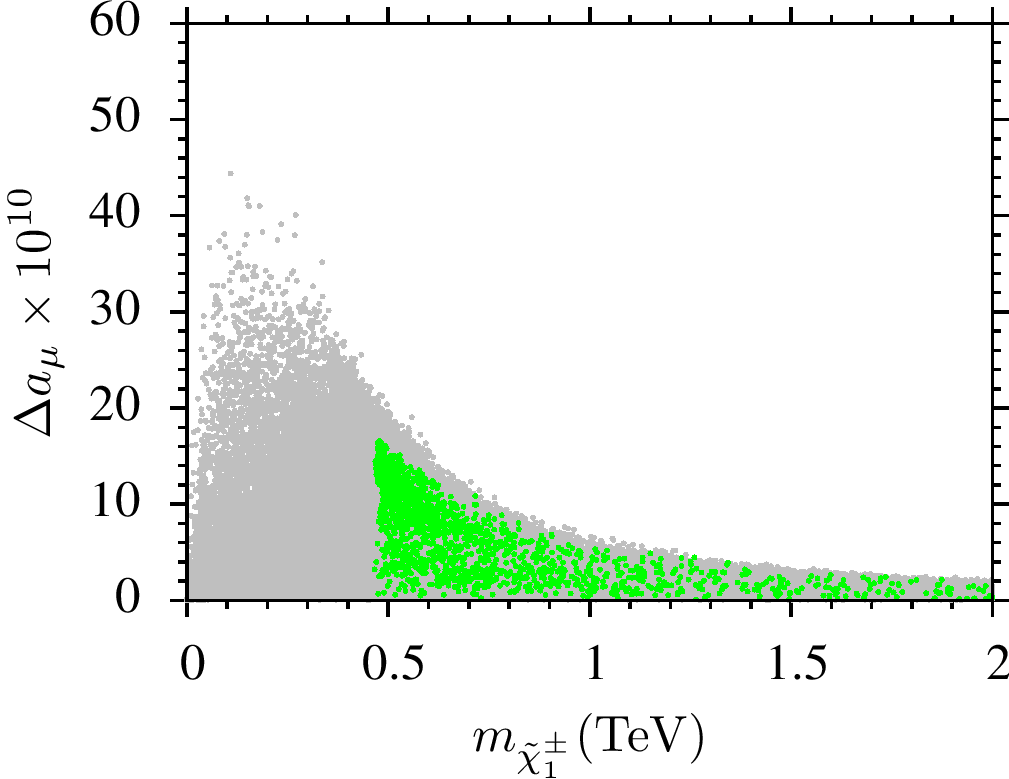}}
\caption{$\Delta a_{\mu}$ as function of $m_{\tilde{\chi}_{1}^{0}}$ (top) and $m_{\tilde{\chi}_{1}^{\pm}}$ (bottom).}
\label{fig2}
\end{figure}

In this regard, the parameters relevant to muon $g-2$ in BLSSM with inverse seesaw can be listed as $M_{\tilde{B}'}$, $m_{\tilde{\nu}^{R}_{1}}$, $\tilde{g}$, and $Y_{\nu}$ in addition to those present in MSSM, where $M_{\tilde{B}'}$ is the soft SUSY breaking mass term for the $\tilde{B}'$, $m_{\tilde{\nu}^{i}_{1}}$ and $m_{\tilde{\nu}^{R}_{1}}$ are the masses of the lightest CP-even and CP-odd sneutrinos respectively. Besides the mass eigenstates, $\tilde{g}$, which quantifies the kinetic gauge mixing between $U(1)_{Y}$ and $U(1)_{B-L}$, and Dirac Yukawa coupling associated with the neutrinos denoted by $Y_{\nu}$ are also effective in supersymmetric contributions to $a_{\mu}$. In scanning the parameter space, we have employed the Metropolis-Hastings algorithm as described in \cite{Belanger:2009ti}, and used SARAH \cite{Staub:2012pb} and SPheno \cite{Porod:2003um} for the numerical results. The data points collected all satisfy the requirement of radiative electroweak symmetry breaking. After collecting the data, we impose the mass bounds on all the particles \cite{Agashe:2014kda}. We have employed the Higgs mass bound as $123~{\rm GeV}\leq m_{h} \leq 127~{\rm GeV}$ \cite{Aad:2012tfa,Chatrchyan:2012xdj}, where we take into account about 2 GeV uncertainty in Higgs boson mass due to the theoretical uncertainties in calculation of the minimum of the scalar potential,  and the experimental uncertainties in $m_{t}$ and $\alpha_{s}$. We also employ the gluino mass bound: $m_{\tilde{g}} \geq 1$ TeV \cite{Aad:2014lra} and the neutral gauge boson $Z'$ mass bound: $M_{Z'} \geq 2.5$ TeV \cite{ATLAS:2013jma}.

We present our results first in Fig. \ref{fig2}, where we plot the muon $g-2$ as function of lightest neutralino (top) and lightest chargino (bottom). The green points satisfy the mass bounds. As can be seen from the panel $\Delta a_{\mu}-m_{\tilde{\chi}_{1}^{0}}$, muon $g-2$ puts a sharp upper bound on the neutralino as $m_{\tilde{\chi}_{1}^{0}}\lesssim 400$ GeV. Similarly, $\Delta a_{\mu}-m_{\tilde{\chi}_{1}^{\pm}}$ exhibits a sharp bound on the lightest chargino mass at about 600 GeV. Also the chargino cannot be lighter than about 500 GeV because of the mass bounds we employed in our analysis. Therefore, the sneutrino-chargino contribution is rather suppressed by the chargino mass. Both planes indicate that the best result for muon $g-2$ is found about $17\times 10^{-10}$ consistently with the constraints.

\begin{figure}[h!]
\centering
\subfigure{\includegraphics[height=6cm,width=7.75cm]{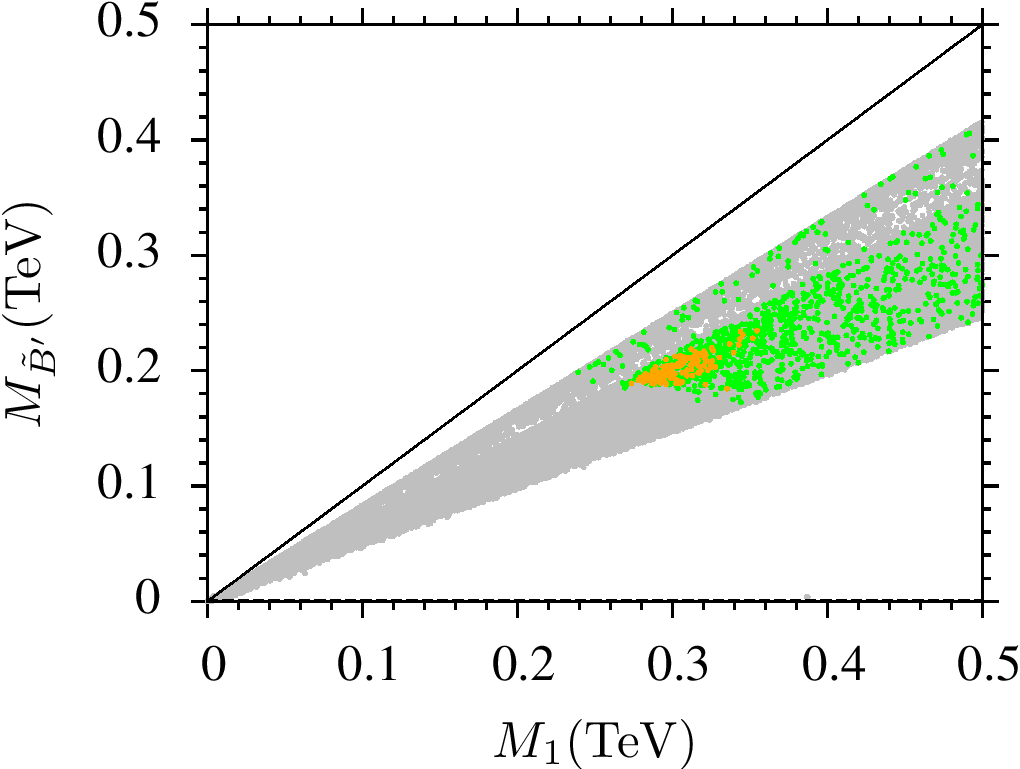}}\\
\subfigure{\includegraphics[height=6cm,width=7.75cm]{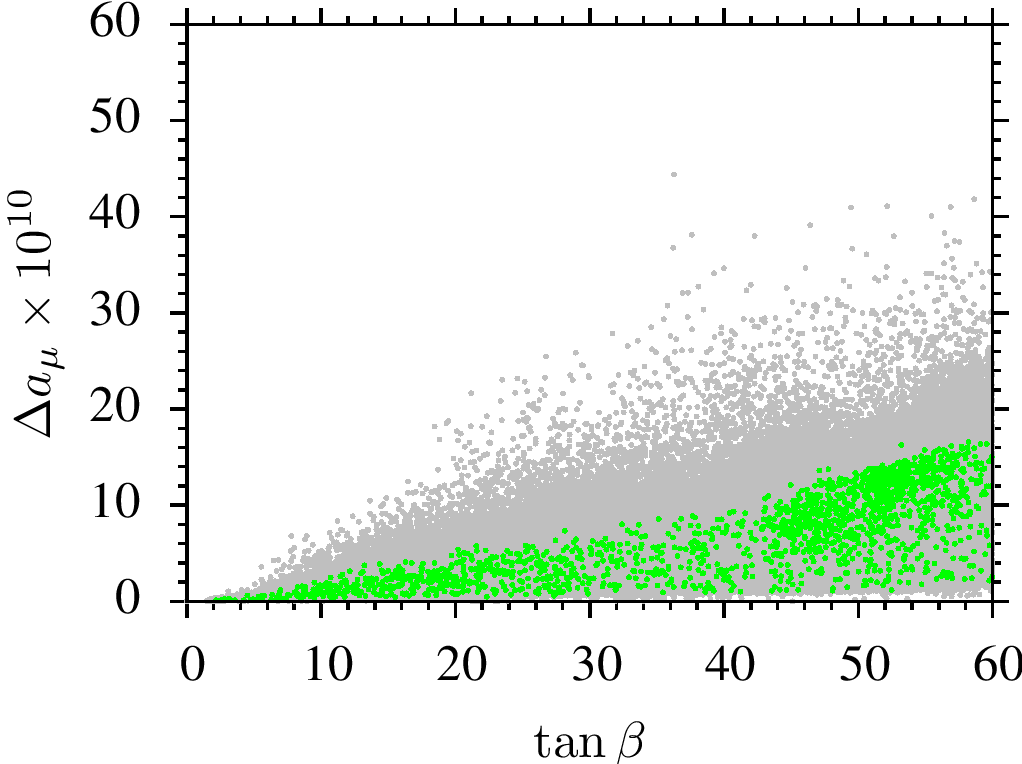}}
\caption{The correlation between the masses of $\tilde{B}'$ and $\tilde{B}$(top) and $\Delta a_{\mu}$ versus $\tan\beta$ (bottom).}
\label{fig4}
\end{figure}

In Fig. \ref{fig4} we display the correlation between the masses of the light neutralinos; $\tilde{B}'$ and $\tilde{B}$ (top) and also the dependence of  $\Delta a_{\mu}$ on $\tan\beta$ (bottom). The color coding is the same as Figure \ref{fig2} except the plane $M_{\tilde{B}'}-M_{1}$ in which the yellow points are a subset of green and they satisfy the muon $g-2$ condition within $2\sigma$. The unit line in this plane indicates the region with $M_{\tilde{B}'}=M_{1}$. As can be seen from the $M_{\tilde{B}'}-M_{1}$ plane, $M_{\tilde{B}'}$ is lighter than $M_{1}$. Hence, Bino and $B'-$ino dominate in muon $g-2$ results through the mixing of neutralinos. According to the results, $M_{\tilde{B}'}$ is bounded in the interval $\sim 200-250$ GeV, while $M_{1}$ in $\sim 250-360$ GeV. One can obtain some significant contributions to muon $g-2$ in such a narrow mass intervals for $\tilde{B}'$ and $\tilde{B}$ with the help of $\tan\beta$ enhancement as shown in the plot of $\Delta a_{\mu}-\tan\beta$ that muon $g-2$ within $2\sigma$ requires $\tan\beta \gtrsim 45$. 

\begin{figure}[h!]
\hspace{-0.7cm}\subfigure{\includegraphics[height=6cm,width=7.75cm]{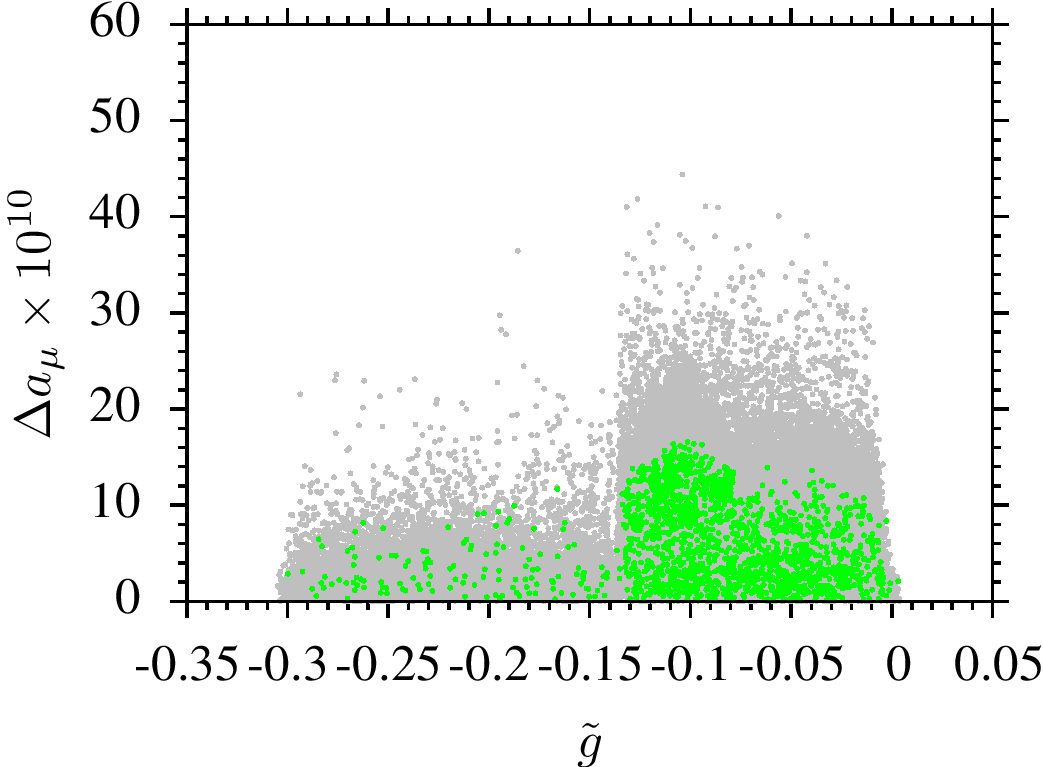}}
\caption{$\Delta a_{\mu}$ as function of $\tilde{g}$. The color coding is the same as Figure \ref{fig2}.}
\label{fig5}
\end{figure}

It is worth noting that the contributions from $M_{\tilde{B}'}$ and $M_{B}$ is further enhanced by their gauge kinetic mixing in smuon-neutralino channel as given in Eqs.(\ref{NLaij}, \ref{NRaij}). In Fig. \ref{fig5} we show the dependence of muon $g-2$ on the gauge kinetic mixing characterized with $\tilde{g}$. The color coding is the same as Figure \ref{fig2}. The figure indicates that $\tilde{g} \sim -0.1$ is  favorably for enhancing $\Delta a_{\mu}$.  

In conclusion, we have found that the supersymmetric contribution to muon $g-2$ in BLSSM with inverse seesaw mostly relies on the light Bino with $M_{1} \gtrsim 250$ GeV, and even lighter $\tilde{B}'$ of mass about 180 GeV. In this case the lightest neutralino mass eigenstate is either mostly $\tilde{B}$ or a linear superposition of $\tilde{B}'$ and $\tilde{B}$. The $\tan\beta$ dependence of $\Delta a_{\mu}$ is also represented, and we have found that $\tan\beta$ needs to be $\gtrsim 45$ in order to raise the supersymmetric contribution such that muon $g-2$ results satisfy the measurement within $2\sigma$. In addition, the kinetic gauge mixing coupling $\tilde{g}$ enhance the supersymmetric contribution to muon $g-2$. We have revealed in our analysis that muon $g-2$ favors $\tilde{g}\approx -0.1$. The gauge kinetic mixing affects the contribution from smuon-neutralino channel, and with its enhancement, the smuons can have the mass about a TeV, and hence the higher $m_{0}$ values can still yield results compatible with muon $g-2$ as well as consistent with the 125 GeV Higgs boson mass.

\textbf{Acknowledgment }

The work of SK is partially supported by the ICTP grant AC-80 and the STDF project 13858.  The work of CS\"{U} is supported in part by the Scientific and Technological Research Council of Turkey (TUBITAK) Grant no. MFAG-114F461 . This work used the Extreme Science and Engineering Discovery Environment (XSEDE), which is supported by the National Science Foundation grant number OCI-1053575.

\end{document}